\documentclass[fp,twocolumn]{jpsj3}
\usepackage{amsmath}
\usepackage{color}

\newcommand{\YbNiAl}{YbNi$_3$Al$_9$}
\newcommand{\YbNiCuAl}{Yb(Ni$_{1-x}$Cu$_x$)$_3$Al$_9$}

\title{Conical Magnetic Structure and Atomic Displacements in Chiral Helimagnet Yb(Ni,Cu)$_3$Al$_9$ in Magnetic Fields along the Helical $c$ Axis}

\author{Takeshi Matsumura$^{1}$, Mitsuru Tsukagoshi$^1$, Shota Nakamura$^2$, and Shigeo Ohara$^2$}
\inst{
$^1$Department of Quantum Matter, ADSE, Hiroshima University, Higashi-Hiroshima 739-8530, Japan \\
$^2$Department of Physical Science and Engineering, Graduate School of Engineering, Nagoya Institute of Technology, Nagoya 466-8555, Japan \\
} 

\abst{
We investigated the conical magnetic state of a uniaxial chiral helimagnet \YbNiCuAl\ induced in magnetic fields applied along the $c$ axis, which coincides with the helical axis at zero field. Using resonant X-ray diffraction, we clearly observed the disappearance of magnetic satellite peaks, corresponding to the transition from the conical to the field-induced ferromagnetic state. 
The critical fields were determined to be 4 T for $x=0$ and 7 T for $x= 0.05$, which were hardly discernible in the magnetization curves. 
We also found that atomic displacements with the same propagation vector emerge simultaneously with the onset of the conical order. 
The transition temperature $T_{\text{N}}$ and the critical fields for $H \parallel c$ ($H_{\text{c}}^{z}$) and $H\perp c$ ($H_{\text{c}}^{x}$) are discussed on the basis of a mean-field calculation for a simple $q=1$ model of the magnetic structure. 
We propose that $T_{\text{N}}$ and $H_{\text{c}}^{z}$ primarily reflect the dominant intralayer exchange interactions within the honeycomb Yb-layer, whereas $H_{\text{c}}^{x}$ is governed by the much weaker interlayer coupling. 
(\today)
}

\begin{document}
\maketitle

\section{Introduction}
In chiral magnets, where the crystal structure possesses neither an inversion center nor a mirror plane, nontrivial spin structures are often realized as a result of Dyzaloshinskii-Moriya (DM)-type antisymmetric exchange interaction. A typical example is the formation of a chiral soliton lattice (CSL) in a uniaxial chiral magnet such as CrNb$_3$S$_6$ \cite{Kishine15,Togawa16,Togawa12,Togawa13,Totawa15}. 
A periodic lattice of twisted spin structures is realized by applying a magnetic field perpendicular to the helical axis. This state is also realized in a rare-earth $4f$ electron system \YbNiCuAl, which exhibits helimagnetic ordering at zero field with magnetic moments lying in the $ab$ plane and propagating along the $c$ axis  \cite{Tobash11,Ohara11,Hirayama12,Yamashita12,Miyazaki12,Ohara14,Matsumura17,Ninomiya18a,Aoki18}. 
Various related studies have been reported in this \textit{R}Ni$_3$\textit{X}$_9$ family of chiral magnets (\textit{R}=rare earth, \textit{X}=Al or Ga), including long-period antiferromagnetic helical ordering in GdNi$_3$Ga$_9$, competition between helimagnetic and ferroquadurpole orderings in DyNi$_3$Ga$_9$, Ising-type antiferromagnetic orderings in ErNi$_3$Al$_9$ and TmNi$_3$Al$_9$, and pressure-induced magnetism in YbNi$_3$Ga$_9$~\cite{Utsumi12,Utsumi20,Wang20a,Wang20b,Shishido21,Silva17,Nakamura23,Ninomiya17,Mendonca18,Ishii18,Ishii19,Tsukagoshi22,Tsukagoshi23b,Ninomiya18b,Ge22,Matsubayashi15,Umeo18,Ota20}. 

The helimagnetic structure of \YbNiAl\ is shown in Fig.~\ref{fig:Escan0T}(a). The two Yb atoms occupying the $6c$ Wyckoff sites of the $R32$ space group, labeled Yb-a and Yb-b, are located at $(0, 0, 0.167)$ and $(0, 0, -0.167)$, respectively. They form honeycomb layers stacked along the $c$ axis with an interlayer spacing of $c/3=9.121$ \AA. In addition, the Yb layers are separated by five Al and two Ni layers \cite{Gladyshevskii93,Nakamura20}. The nearest neighbor distance between Yb-a and Yb-b atoms within a layer is $a/\sqrt{3}=4.199$ \AA, which suggests that the intralayer exchange interactions may be stronger than the interlayer interactions. 

The magnetic structure of \YbNiAl, characterized by the propagation vector $\mib{q}$=(0, 0, 0.82), has a unique helicity that has a one-to-one correspondence with the crystal chirality. 
In the right-handed crystal, the magnetic moments rotate clockwise by $98.4^{\circ}$ when propagating through one layer along the $c$ axis. The magnetic moment of Yb-a on the layer at $z=z_j$ is expressed as $\mib{\mu}_{\text{a},j} \propto \hat{\mib{x}}\cos qz_j  - \hat{\mib{y}} \sin qz_j$ \cite{Matsumura17}. 
The angle between the magnetic moments of Yb-a and Yb-b within a layer has been reported to be approximately $20.5^{\circ}$ \cite{Ninomiya18a}. 

In \YbNiAl\ with $T_{\text{N}}$=3.4 K, when a magnetic field is applied perpendicular to the $c$ axis, the helimagnetic order easily transforms into a ferromagnetic state at a critical field of $H_{\text{c}}^x$=1 kG \cite{Yamashita12}. 
In \YbNiCuAl\ with $x$=0.06 and $T_{\text{N}}$=6.5 K, for $H\perp c$, a CSL state develops with increasing field, and the system eventually becomes ferromagnetic at $H_{\text{c}}^x$=10 kG \cite{Ohara14,Matsumura17}. Compared with \YbNiAl, $H_{\text{c}}^x$ increases by an order of magnitude, whereas $T_{\text{N}}$ increases only by a factor of two, which remains an unresolved issue. 
When the magnetic field is applied parallel to the helical $c$ axis, a conical order is expected to appear, accompanied by a uniform ferromagnetic component along the $c$ axis while maintaining the helical arrangement of the in-plane components. 
With further increasing field strength, a critical field $H_{\text{c}}^z$ should occur, above which the helical component vanishes and only the ferromagnetic component remains. However, no anomaly has been detected in the magnetization process along the $c$ axis \cite{Yamashita12,Ito20}. 

The purpose of this study is to investigate the conical magnetic structures in \YbNiAl\ and \YbNiCuAl\ with $x$=0.05 for $H \parallel c$ using resonant X-ray diffraction. 
We show that the critical fields are 4 T for $x=0$ and 7 T for $x= 0.05$. 
We then discuss the relationships among $T_{\text{N}}$, $H_{\text{c}}^z$, and $H_{\text{c}}^x$ for these compounds on the basis of a mean-field calculation.  
We propose that $T_{\text{N}}$ and $H_{\text{c}}^{z}$ primarily reflect the dominant exchange interaction within the honeycomb Yb layer, whereas $H_{\text{c}}^{x}$ is governed by much weaker interlayer coupling.  
Field-induced atomic displacements accompanying the conical magnetic order are also presented. 

\section{Experiment}
\begin{figure}[t]
\begin{center}
\includegraphics[width=8cm]{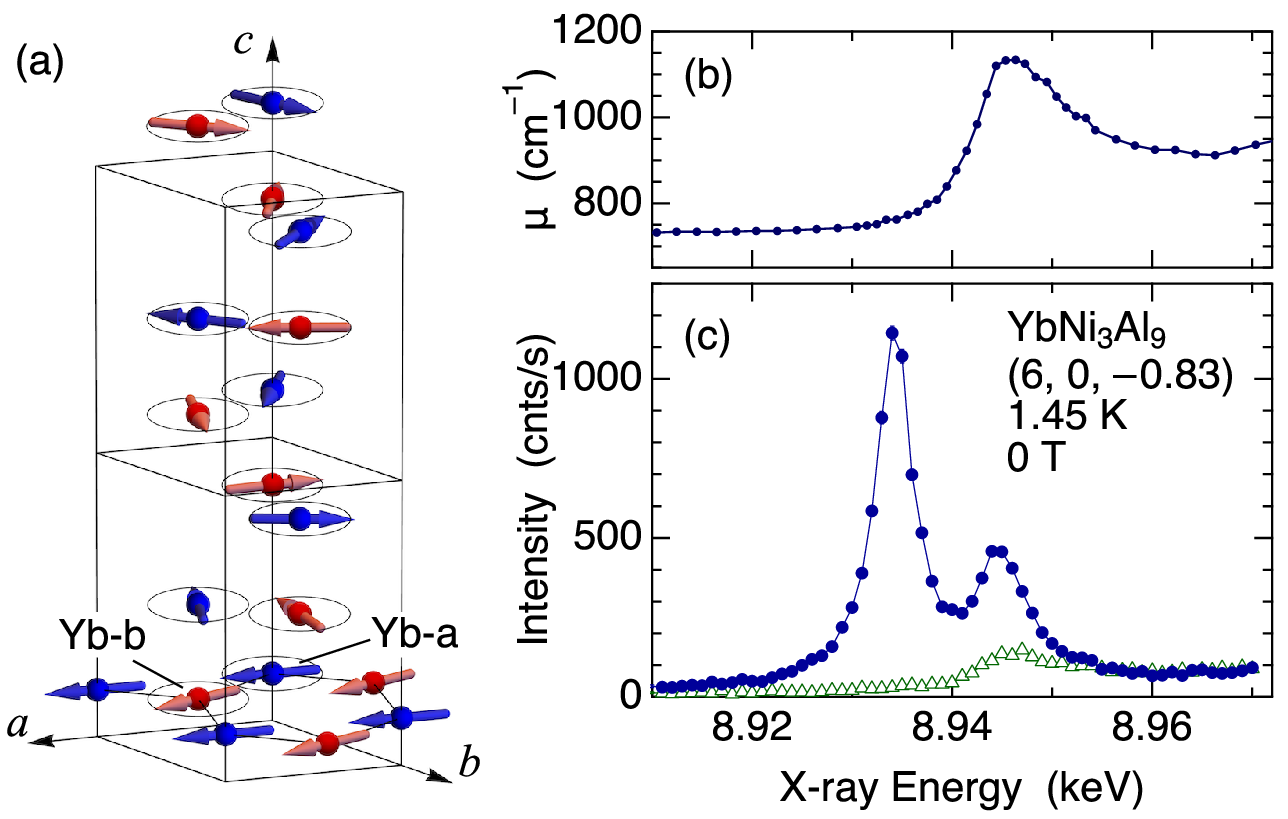}
\end{center}
\caption{(Color online) 
(a) Magnetic structure of \YbNiAl\ for the right-handed crystal. The magnetic moments of Yb-a and Yb-b are indicated by different colors. The magnetic moments rotate clockwise when propagating along the $c$ axis. 
(b) Absorption coefficient of \YbNiAl\ deduced from the fluorescence spectrum. 
(c) X-ray energy dependence of the intensity of the (6, 0, $-0.83$) resonant magnetic Bragg diffraction. Triangles represent the background intensities. 
}
\label{fig:Escan0T}
\end{figure}

Resonant X-ray diffraction experiments were performed at the RIKEN beamline BL19LXU of SPring-8 (for the $x=0$ sample) and at beamline 3A of Photon Factory, KEK (for the $x=0.05$ sample)~\cite{Yabashi01}. 
Single crystalline samples of \YbNiAl\ and \YbNiCuAl, with mirror-polished surfaces normal to the $a^*$-axis, were mounted in an 8 Tesla vertical-field cryomagnet with the $c$ axis aligned vertically. The magnetic Bragg reflection at (6, 0, $q$) was investigated by tilting the $\chi$ axis of the diffractometer.  The X-ray energy was tuned around the Yb $L_3$ absorption edge. 

A diamond phase retarder system was employed to convert the incident linearly polarized X-ray beam into right-handed circular polarization (RCP) or left-handed circular polarization (LCP). 
By rotating the diamond phase plate around the Bragg angle $\theta_{\text{B}}$ of 220 reflection (at BL19LXU), i.e., by tuning $\Delta\theta_{\text{PR}} = \theta_{\text{PR}} - \theta_{\text{B}}$, a circularly polarized beam was obtained \cite{Matsumura17}. 
Using the Stokes parameters, the degrees of circular and linear polarization in our setup are expressed as $P_2 = -\sin (\gamma/\Delta\theta_{\text{PR}})$ ($+1$ for RCP and $-1$ for LCP) and $P_3 = -\cos (\gamma/\Delta\theta_{\text{PR}})$ ($+1$ for $\sigma$ and $-1$ for $\pi$), where $\gamma$ is an experimentally determined parameter of the phase plate obtained from the $\Delta\theta_{\text{PR}}$ dependence of the (6 0 0) fundamental reflection. The obtained value was $\gamma=0.0305^{\circ}$ (see Appendix). The degree of $45^{\circ}$ linear polarization, $P_1$, is zero in the present geometry. 

Figure \ref{fig:Escan0T}(b) shows the absorption coefficient of \YbNiAl\ deduced from the fluorescence spectrum. 
The absorption coefficent $\mu$ was converted to the imaginary part of the atomic scattering factor $f^{\prime\prime}$ of Yb, from which the real part $f^{\prime}$ was obtained by a Kramers-Kronig transformation. 
These scattering factors near the absorption edge were used to calculate the energy dependence of the (4 1 0) and (5 $\bar{1}$ 0) fundamental Bragg reflections and to compare the results with the experimental observations. From this analysis, the crystal chirality of the \YbNiAl\ sample used in the present experiment was determined to be right handed (see Appendix) \cite{Matsumura17}.

\section{Results and Analysis}
\subsection{\YbNiAl}
Figure \ref{fig:Escan0T}(c) shows the energy dependence of the intensity of the helimagnetic Bragg peak, which exhibits resonant enhancements at 8.934 keV and 8.944 keV, corresponding to the $E2$ and $E1$ resonances, respectively. Strong enhancement is observed at the $E2$ resonance compared with the previous measurements performed along the (0, 0, $L$) direction using the sample with an $ab$-place surface. 
This is due to the difference of geometrical structure factor for the $T_{1u}$-type magnetic octupole moment contributing to the $E2$ resonance \cite{Matsumura17}. 
n the present geometry for the (6, 0, $q$) reflection, we expect an intensity three times larger than that for, for example, the (0, 0, 27+$q$) reflection \cite{Nagao06}. 

\begin{figure}[t]
\begin{center}
\includegraphics[width=8cm]{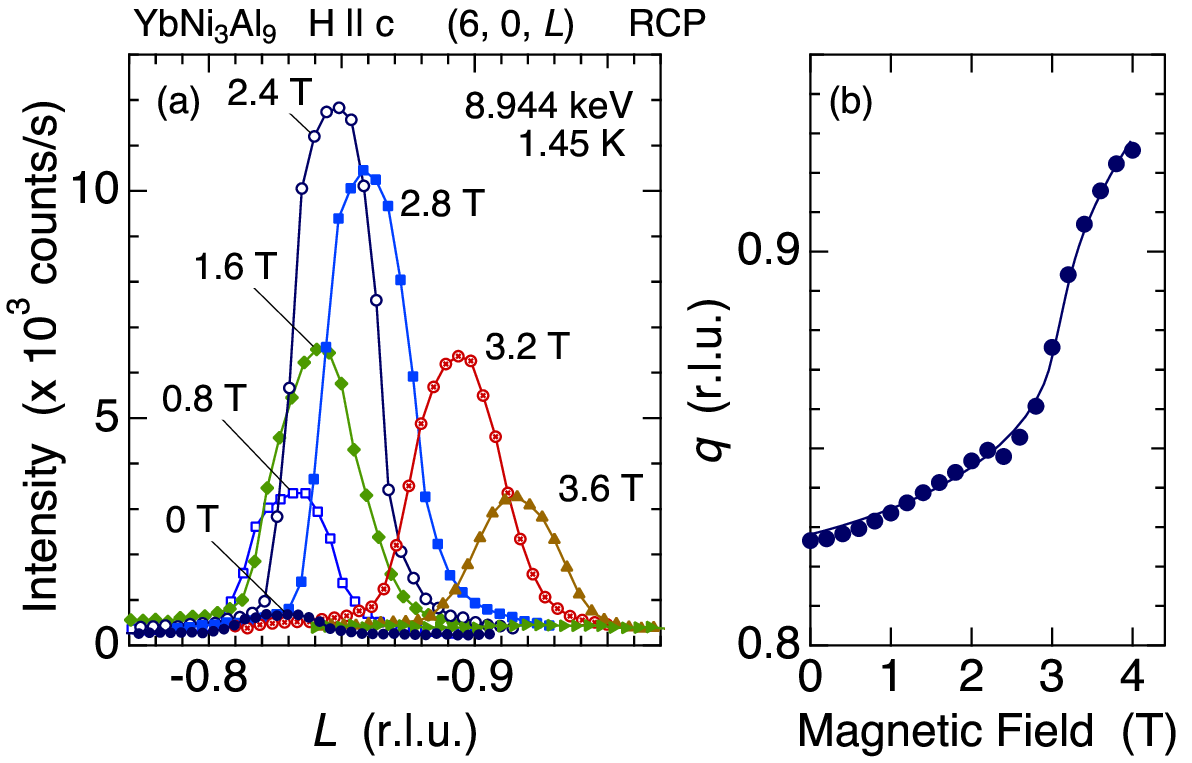}
\end{center}
\caption{(Color online) 
(a) Reciprocal-space scans along (6, 0, $L$) at several magnetic fields applied along the helical $c$ axis. The incident X-ray polarization is RCP. 
(b) Magnetic-field dependence of the wave number $q$. 
}
\label{fig:Lscan0}
\end{figure}

Figure \ref{fig:Lscan0}(a) shows the magnetic-field dependence of the peak profile at the resonance energy of 8.944 keV. 
The intensity increases markedly with increasing field, reaches a maximum around 2.4 T, and then decreases and eventually disappears at 4 T. The field dependence of the wavenumber $q$, which increases with increasing field, is plotted in Fig.~\ref{fig:Lscan0}(b). 
As discussed below, the pronounced enhancement of the intensity originates from the nonresonant Thomson scattering arising from field-induced atomic displacements. 
No change in the lattice parameter was observed beyond the experimental accuracy of $\sim 3\times 10^{-4}$, as was also the case for $x=0.05$.

\begin{figure}[t]
\begin{center}
\includegraphics[width=8.2cm]{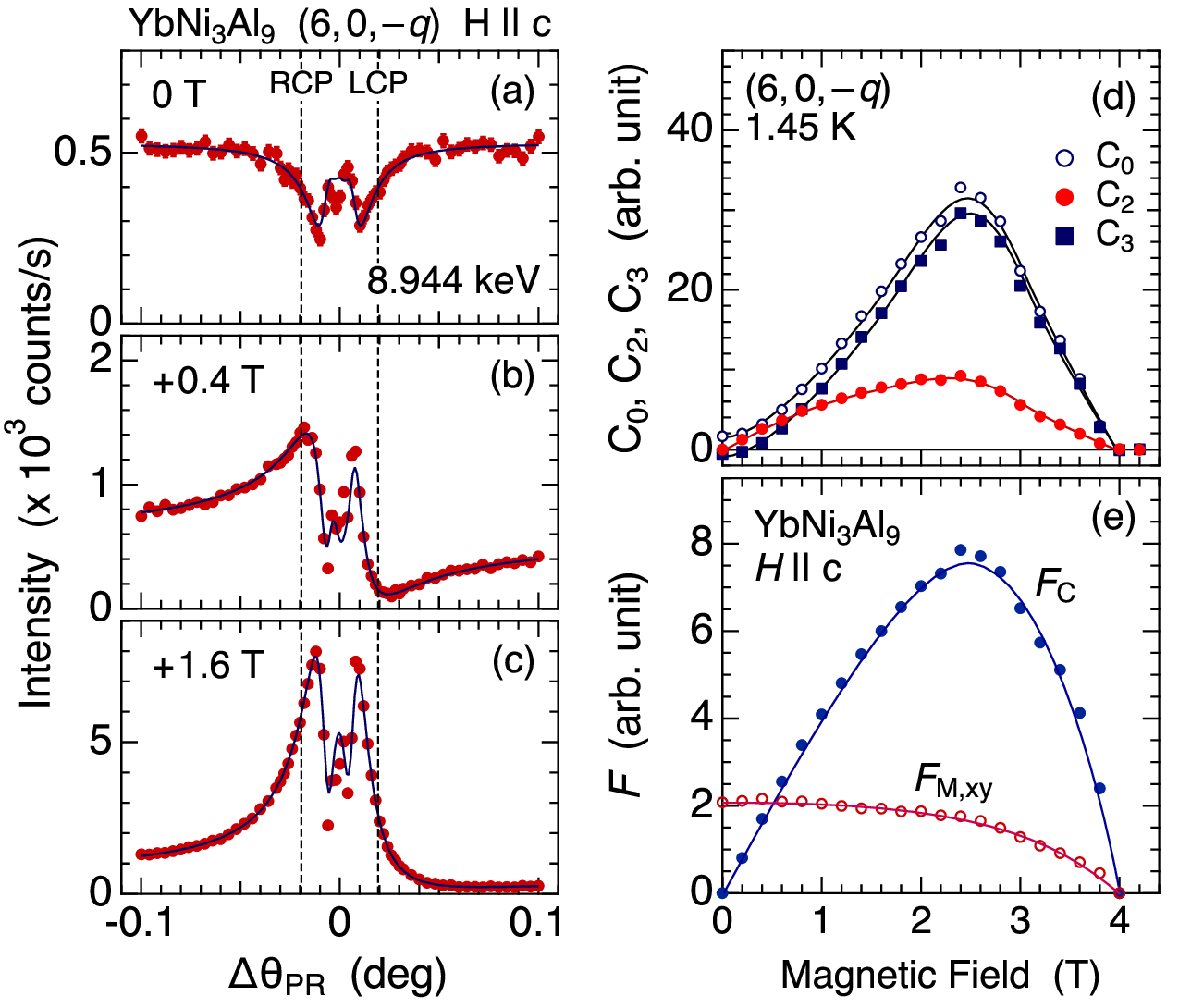}
\end{center}
\caption{(Color online) 
(a--c) Incident-polarization ($\Delta\theta_{\text{PR}}$) dependence of the peak intensity in magnetic fields of 0, 0.4, and 1.6 T applied along the $c$ axis, without polarization analysis. 
The background has been subtracted. The vertical dashed lines indicate the positions of the RCP and LCP states. 
The solid lines represent fits obtained by convolving Eq.~(\ref{eq:CrossSec2}) with a Gaussian resolution function.} 
(d) Magnetic-field dependence of the parameters $C_0$, $C_2$, and $C_3$ obtained from the $\Delta\theta_{\text{PR}}$ scans. 
(e) Magnetic-field dependences of the crystal and magnetic structure factors, $F_{\text{C}}$ and $F_{\text{M},xy}$, representing the atomic displacement and the in-plane helical component of the conical magnetic structure, respectively, with the same propagation vector $(0, 0, q)$. 
\label{fig:PRthscans0}
\end{figure}


Figures \ref{fig:PRthscans0}(a)--(c) show the incident polarization dependence of the peak intensity. 
The scattering intensity can be expressed by
\begin{align}
I &= 
\frac{1}{2} \bigl(\, |F_{\sigma\sigma'}|^2 + |F_{\sigma\pi'}|^2 + |F_{\pi\sigma'}|^2 + |F_{\pi\pi'}|^2 \,\bigr) \nonumber \\
 &\;\;\;\; + P_1 \text{Re} \bigl\{\, F_{\pi\sigma'}^*F_{\sigma\sigma'} + F_{\pi\pi'}^*F_{\sigma\pi'} \,\bigr\} \nonumber \\
 &\;\;\;\; + P_2 \text{Im} \bigl\{\, F_{\pi\sigma'}^*F_{\sigma\sigma'} + F_{\pi\pi'}^*F_{\sigma\pi'} \,\bigr\} 
 \label{eq:CrossSec1} \\
 &\;\;\;\; +  \frac{1}{2} P_3\bigl(\, |F_{\sigma\sigma'}|^2 + |F_{\sigma\pi'}|^2 - |F_{\pi\sigma'}|^2 - |F_{\pi\pi'}|^2 \,\bigr) 
 \,, \nonumber
\end{align}
using the four scattering amplitudes for the $\sigma$-$\sigma'$, $\sigma$-$\pi'$, $\pi$-$\sigma'$, and $\pi$-$\pi'$ polarization channels. 
Therefore, the intensity for an incident beam described by $(P_1, P_2, P_3)$ can generally be written as
\begin{equation}
I = C_0 + C_1 P_1 + C_2 P_2 + C_3 P_3 \,.
\label{eq:CrossSec2}
\end{equation}
This expression is used as the fitting function for the $\Delta\theta_{\text{PR}}$ scan with three parameters of $C_0$, $C_2$, and $C_3$. 
In our setup, $P_1=0$. 
 
The solid lines in Figs.~\ref{fig:PRthscans0}(a-c) represent fits obtained by convolving the above equation with a Gaussian resolution function.
The parameters obtained from these fits were normalized by the intensity of the (6 0 0) fundamental reflection, and are summarized in Fig.~\ref{fig:PRthscans0}(d) in arbitrary units.

The reason for the nearly equal values of $C_0$ and $C_3$ at high fields in Fig.~\ref{fig:PRthscans0}(d) is that most of the intensity originates from $F_{\sigma\sigma'}$, reflecting the field-induced Thomson scattering. 
The asymmetric $\Delta\theta_{\text{PR}}$ dependence observed in Fig.~\ref{fig:PRthscans0}(b) and (c) also arises mainly from the term $F_{\pi\sigma'}^*F_{\sigma\sigma'}$. 
It should be noted here that this scattering geometry, where the scattering plane is almost parallel to the helical plane, is not sensitive to the magnetic helicity. Since we use a right-handed crystal, the magnetic structure factor for $\mib{Q}=(6, 0, q)$ is expressed as $\hat{\mib{x}}+i\hat{\mib{y}}$ \cite{Matsumura17}. 
In this geometry, the magnetic scattering amplitude for the $E1$ dipole transition consists only of $F_{\pi\sigma'}$ and $F_{\sigma\pi'}$. 
Because $F_{\pi\sigma'}$ contains the imaginary magnetic structure factor multiplied by the resonant spectral function, and the phase of the spectral function is unknown, we cannot determine the phase of the magnetic structure factor independently of that of the spectral function. 

To analyze the $C_n$ $(n=0, 2, 3)$ parameters shown in Fig.~\ref{fig:PRthscans0}(d), we assume the total structure factor under magnetic fields to be
\begin{align}
F&=F_{\text{C}}(\mib{\varepsilon}^{\prime}\cdot\mib{\varepsilon}) 
+ F_{\text{M},xy}e^{i\phi_1} (\hat{\mib{x}} + i\hat{\mib{y}}) \cdot   (\mib{\varepsilon}^{\prime}\times \mib{\varepsilon} ) 
\;.
\label{eq:FMmodel}
\end{align}
Here, $\mib{\varepsilon}^{\prime}\times \mib{\varepsilon}$ represents the geometrical factor for the $E1$ resonance arising from the magnetic dipole, and $\phi_1$ denotes the phase of the spectral function at the $E1$ resonance energy of 8.944 keV~\cite{Hannon88,Nagao06}. 


Consequently, three parameters remain to determine the intensity: the charge structure factor $F_{\text{C}}$ for nonresonant Thomson scattering, the magnetic structure factor $F_{\text{M},xy}$ reflecting the in-plane component of the conical magnetic structure, and the phase factor $\phi_1$. 
These parameters can be determined from $C_0$, $C_2$, and $C_3$ by numerically solving the equation. 
In this process, $\phi_1$ was fixed to the average value of $-0.55\pi$. 
The resultant charge and magnetic structure factors are shown in Fig.~\ref{fig:PRthscans0}(e). 
$F_{\text{C}}$ increases linearly with magnetic field, reaches a maximum around 2.5 T, and decreases to zero at the critical field of 4 T. 
The in-plane magnetic component exhibits only a slight decrease up to 2.5 T and then smoothly decreases to zero at 4 T. 
These results indicate that the decrease of $F_{\text{C}}$ above 2.5 T is associated with the closing of the magnetic cone. 
As discussed later, this behavior can be interpreted as being related to the $O_{yz}$ and $O_{zx}$ quadrupole moments of Yb, which increase with magnetic field as long as the in-plane component remains nearly constant up to 2.5 T, but begin to decrease simultaneously with the closing of the cone.

At 0 T [Fig.~\ref{fig:PRthscans0}(a)], the weaker intensity at $\Delta\theta_{\text{PR}}=\pm 0.01^{\circ}$ for $\sigma$ incident polarization corresponds to a negative value of $C_3$. 
If we assume a helical magnetic order lying in the $ab$ plane, the $E1$ resonance scattering factor arising from the magnetic dipole term, represented by $(\mib{\varepsilon}^{\prime} \times \mib{\varepsilon}) \cdot \mib{m}$ in the $(6, 0, -q)$ reflection geometry, yields $F_{\sigma\sigma'}=0$, $F_{\pi\pi'}\approx 0$ and $|F_{\sigma\pi'}| \simeq |F_{\pi\sigma'}|$, and therefore $C_3 \approx 0$. 
The observed negative value of $C_3$ indicates $|F_{\sigma\pi'}| < |F_{\pi\sigma'}|$. This cannot be explained by the $E1$ resonance alone, and it is necessary to take into account a superposition with the tail of the $E2$ resonance from a $T_{1u}$-type magnetic octupole (see Appendix), as reported previously~\cite{Matsumura17}. 
The octupole contribution to the intensity, however, is only discernible at 0 T. At higher magnetic fields, where the signal from Thomson scattering dominates, it can be neglected.

When the field direction is reversed, the asymmetries in Figs.~\ref{fig:PRthscans0}(b) and (c) are also reversed, indicating that the sign of $C_2$ is inverted. This behavior arises from the change in the sign of $F_{\sigma\sigma'}$ upon field reversal. The magnetic part of the structure factor does not change, meaning that the magnetic helicity remains unchanged. In other words, when the field direction is reversed, $F_{\text{C}}$ changes sign, whereas $F_{\text{M},xy}$ is symmetric with the field direction (see Appendix). 

\begin{figure}[t]
\begin{center}
\includegraphics[width=8cm]{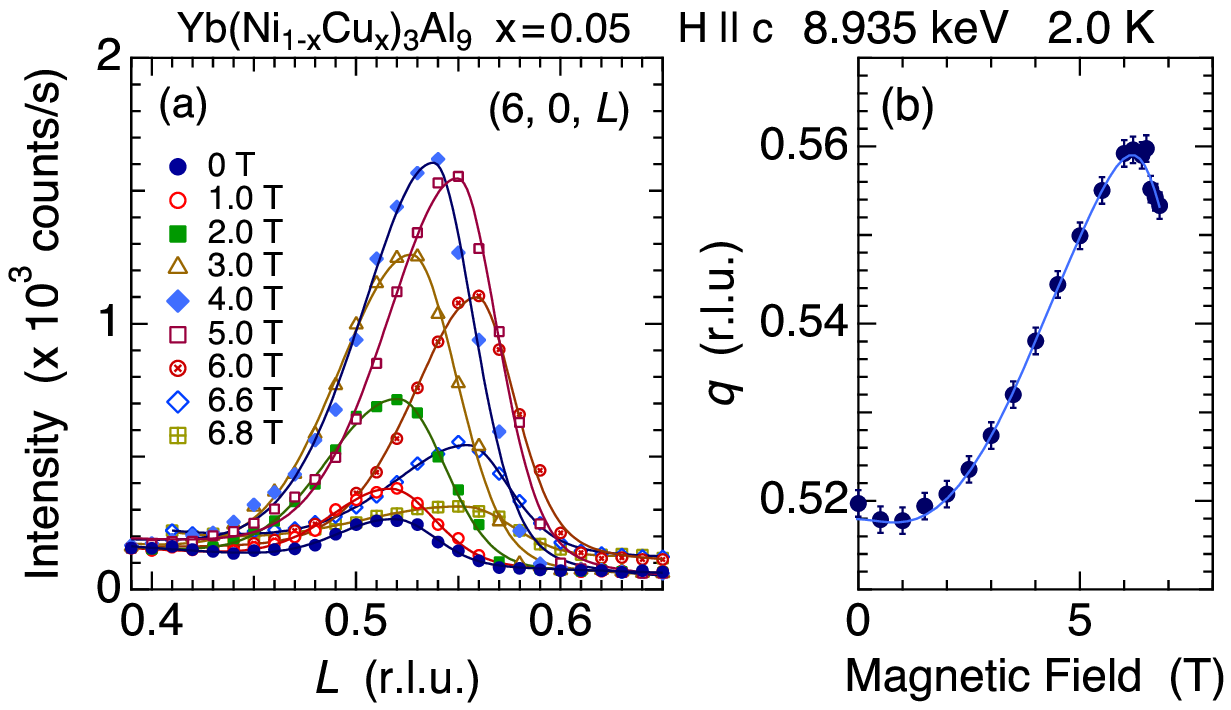}
\end{center}
\caption{(Color online) 
(a) Reciprocal-space scan along $(6, 0, L)$ in magnetic fields applied along the $c$ axis. The incident X-rays are $\pi$ polarized ($P_3 = -1$). 
(b) Magnetic-field dependence of the wavenumber $q$. 
}
\label{fig:Lscan006}
\end{figure}

\subsection{\YbNiCuAl\ : $x$=0.05}
Figure \ref{fig:Lscan006}(a) shows the magnetic-field dependence of the peak profile at the $E2$ resonance energy of 8.935 keV. 
From the peak position of $L=0.52$ at 0 T,  the Cu concentration of this sample is estimated to be $x=0.05$ \cite{Matsumura17}. 
The magnetic-field dependence of the wavenumber $q$ is plotted in Fig. \ref{fig:Lscan006}(b). 
As in the case of $x=0$, the intensity increases markedly with increasing field, reflecting the appearance of nonresonant Thomson scattering due to atomic displacements, and the wavenumber also increases with increasing field. 
The intensity finally disappears at the critical field of 7 T. 

\begin{figure}[t]
\begin{center}
\includegraphics[width=8cm]{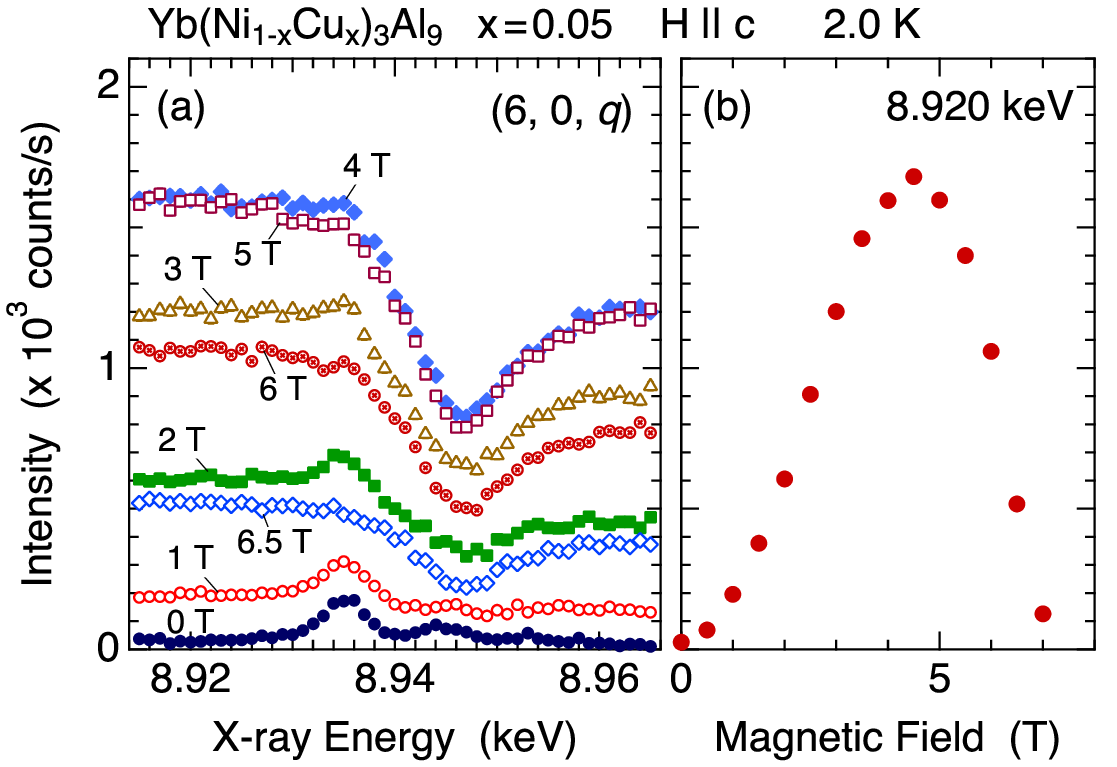}
\end{center}
\caption{(Color online) 
(a) X-ray energy dependence of the peak intensity in magnetic fields applied along the $c$ axis. 
(b) Magnetic-field dependence of the nonresonant intensity at 8.920 keV. 
}
\label{fig:Escan006}
\end{figure}

Figure \ref{fig:Escan006}(a) shows the X-ray energy dependence of the peak intensity at several magnetic fields. 
At 0 T, the energy dependence exhibits two resonance peaks at 8.935 keV and 8.944 keV, corresponding to the $E2$ and $E1$ resonances, respectively. 
With increasing field, the nonresonant intensity increases, eventually overwhelming the resonant contributions, and the $E2$ resonance peak at 8.935 keV becomes unrecognizable above 4 T. 
As shown in Fig. \ref{fig:Escan006}(b), the nonresonant intensity reaches a maximum at 4.5 T, and decreases to zero at the critical field of 7 T. 
Although not shown explicitly, the field dependence of $F_{\text{C}}$, obtained by taking the square root of the data in Fig. \ref{fig:Escan006}(b), exhibits nearly the same behavior as that for $x=0$ shown in Fig.~\ref{fig:PRthscans0}(e). 
From these observations, we conclude that the ratios between $T_{\text{N}}$ and $H_{\text{c}}^z$ -- (3 K, 4 T) for $x=0$ and (6 K, 7 T) for $x=0.05$ -- are nearly equal, indicating that these quantities reflect the same exchange energy.

\section{Discussions}
\subsection{Mean-field calculation}
\begin{figure}[t]
\begin{center}
\includegraphics[width=8.5cm]{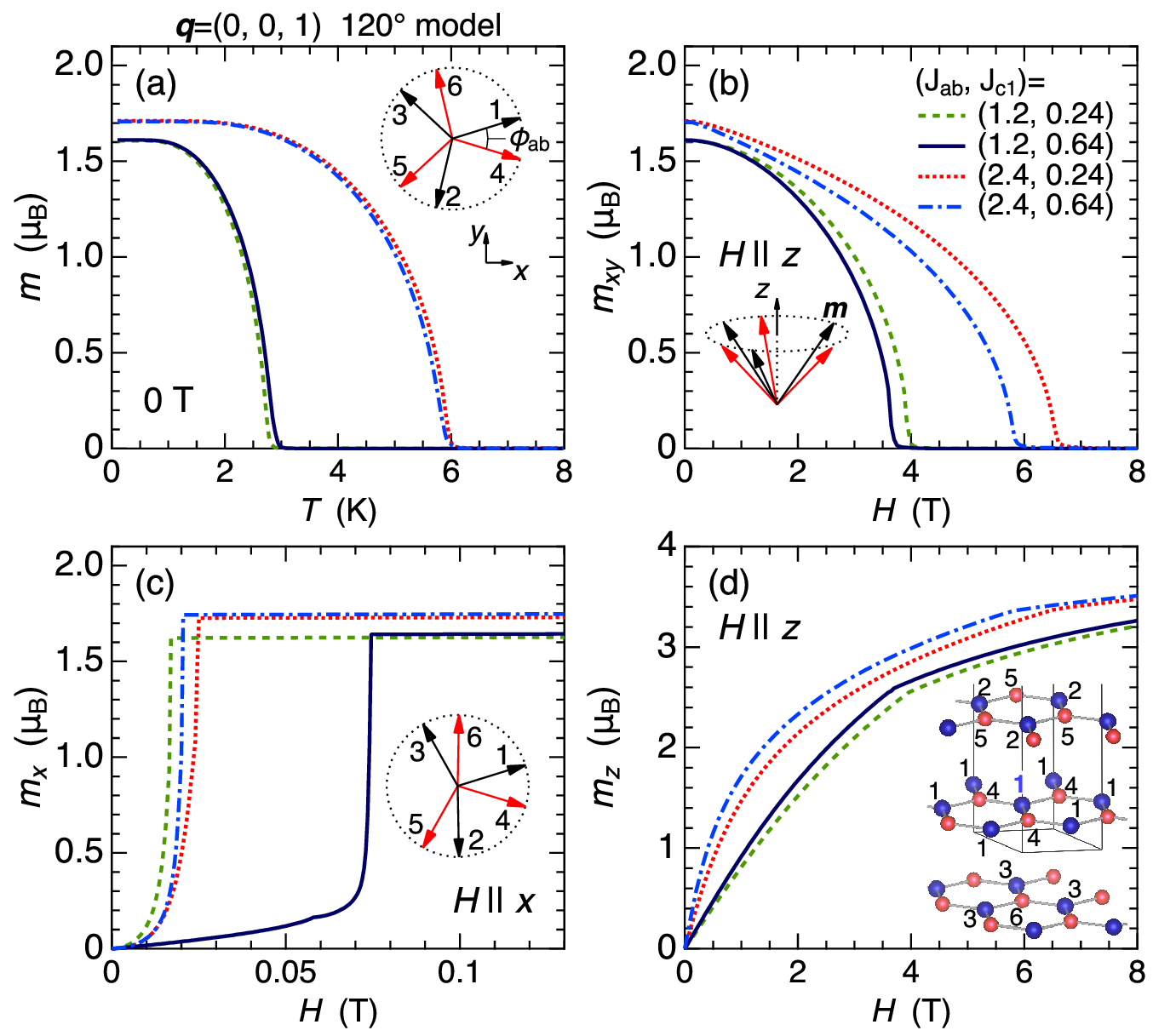}
\end{center}
\caption{(Color online) 
Results of a mean-field calculation for a $120^{\circ}$ model with $\mib{q}=(0, 0, 1)$, consisting of six Yb atoms in a unit cell. 
(a) Temperature dependences of the ordered moment at zero field. 
(b) Magnetic-field dependence of the in-plane component of the conical structure for $H \parallel z$. 
(c) Magnetic-field dependence of the total magnetization for $H \parallel x$. 
(d) Magnetic-field dependence of the total magnetization for $H \parallel z$. The inset shows the numbering configuration of the Yb atoms. The atoms 1--3 and 4--6 belong to the Yb-a and Yb-b sublattices, respectively. 
}
\label{fig:MFcalc}
\end{figure}

To discuss the relation between the N\'eel temperature $T_{\text{N}}$ and the critical field $H_{\text{c}}^z$ for $H \parallel z$ in a simple manner, we introduce a mean-field model assuming a $120^{\circ}$ spin structure with $\mib{q}=(0, 0, 1)$, in which the angle between the Yb-a (or Yb-b) magnetic moments on neighboring layers is $120^{\circ}$ at zero field, as shown in the inset of Fig.~\ref{fig:MFcalc}(a). 
Although $H_{\text{c}}^x$ for $H \parallel x$ is also calculated, it is not our purpose here to discuss the CSL state realized in $H \parallel x$. 
This calculation shows that $T_{\text{N}}$ and $H_{\text{c}}^z$ are proportional to each other and reflect the strong exchange interactions between the nearest and next-nearest neighbor Yb atoms within the honeycomb layer. 
In contrast, $H_{\text{c}}^x$ is determined by the interplay between the strong intralayer and weak interlayer exchange interactions. 
$H_{\text{c}}^x$ is proportional to the interlayer exchange when the intralayer exchange interaction is relatively weak. The physics of helical magnetism in this system is primarily governed by the interlayer exchange and the DM-type exchange interactions. 
Although clarifying this mechanism is the ultimate goal of our study, it is beyond the scope of the present calculation. 

For the Yb-1 atoms shown in the inset of Fig.~\ref{fig:MFcalc}(d), we assume the following mean-field Hamiltonian:  
\begin{align}
\mathcal{H}_1 &= \mathcal{H}_{\text{CEF}} + g\mu_{\text{B}}\mib{J}\cdot\mib{H} 
- J_{\text{ab}}\mib{J} \cdot \langle \mib{J}_4 \rangle - J_{\text{aa}}\mib{J} \cdot \langle \mib{J}_1 \rangle \nonumber \\
 &  
 - J_{\text{c1}}\mib{J} \cdot \langle \mib{J}_6 \rangle - J_{\text{c2}}\mib{J} \cdot ( \langle \mib{J}_2 \rangle + \langle \mib{J}_3 \rangle + \langle \mib{J}_5\rangle) 
\end{align}
Here, $\mathcal{H}_{\text{CEF}}$ denotes the crystalline electric field (CEF) determined from inelastic neutron scattering \cite{Tsukagoshi23}. 
$J_{\text{ab}}=J_{14}$ and $J_{\text{aa}}=J_{11}$ represent the nearest- and next-nearest-neighbor intralayer exchange interactions, respectively, while $J_{\text{c1}}=J_{16}$ and $J_{\text{c2}}=J_{12}=J_{13}=J_{15}$ correspond to the interlayer exchange interactions. 
Although $J_{15}$ is not exactly equal to $J_{12}$, we assume them to be equal for simplicity. 
The numbers of neighboring atoms contributing to $J_{\text{ab}}$, $J_{\text{aa}}$, $J_{\text{c1}}$, and $J_{\text{c2}}$ are three, six, one, and nine, respectively. 
Furthermore, we fix $J_{\text{ab}}=4J_{\text{aa}}$ and $J_{\text{c1}}=-8J_{\text{c2}}$. 
In this model, neglecting the interactions between second-nearest layers, it is necessary to introduce a competition between $J_{\text{c1}}$ and $J_{\text{c2}}$ in order to realize a helical order. 
We take $J_{\text{c1}}$ and $J_{\text{c2}}$ to be ferromagnetic and antiferromagnetic, respectively. 

At high magnetic fields for $H \parallel c$, where the CEF excited states become involved, the exchange interaction needs to be modified due to the orbital-dependent exchange mechanism \cite{Tsukagoshi23}. However, for simplicity, we assume that the exchange interaction between the $z$ components of the magnetic moments is 0.6 times weaker, which yields an $H_{\text{c}}^z$ value roughly consistent with the experiment. 

The results of the calculation are summarized in Fig.~\ref{fig:MFcalc}. 
First, Fig.~\ref{fig:MFcalc}(a) shows that $T_{\text{N}}$ is primarily governed by the intralayer exchange interactions. 
$T_{\text{N}}$ is proportional to $J_{\text{ab}}$ and $J_{\text{aa}}$, in agreement with the theoretical study reported in Ref.~\citen{Shinozaki16}. 
The interlayer interactions have little effect on $T_{\text{N}}$. 
Second, as shown in Fig.~\ref{fig:MFcalc}(b), the critical field $H_{\text{c}}^z$ for $H \parallel c$ is roughly proportional to $T_{\text{N}}$, indicating that $H_{\text{c}}^z$ also reflects the strong intralayer interactions and thereby explains our experimental results. 
$H_{\text{c}}^z$ is mainly proportional to $J_{\text{ab}}$ and $J_{\text{aa}}$, although it is slightly reduced with increasing ferromagnetic interlayer exchange $J_{\text{c1}}$. 

In contrast to $T_{\text{N}}$ and $H_{\text{c}}^z$, the critical field $H_{\text{c}}^x$ is sensitive to the interlayer exchange interactions, which are weak but play a key role in stabilizing the helical magnetism. As shown in Fig.~\ref{fig:MFcalc}(c), $H_{\text{c}}^x$ is enhanced by increasing $J_{\text{c1}}$ and $J_{\text{c2}}$ when $J_{\text{ab}}$ and $J_{\text{aa}}$ are relatively small. However, detailed discussion of $H_{\text{c}}^x$ within this simplified model is not meaningful; as our primary focus here is not on $H_{\text{c}}^x$. 

A more fundamental problem for $H \parallel x$ lies in the relationship between the helical wavenumber $q$ and the critical field \cite{Matsumura17}, which is not addressed here in detail because the wavenumber $q$ is beyond the scope of  the present $120^{\circ}$ model. 
The key question is why $H_{\text{c}}^x$ increases as the wavenumber $q$ decreases, even though a smaller $q$ value implies that the exchange interaction more strongly favors the ferromagnetic state. 
When the interlayer exchange interactions are modified to favor a smaller $q$ value, the critical field  $H_{\text{c}}^x$ necessarily decreases. In contrast, experimentally, $H_{\text{c}}^x$ for $x=0.06$ with $q=0.42$ is 1 T, which is about ten times larger than 0.1 T for $x=0$ with $q=0.82$. 
This behavior is difficult to explain solely in terms of exchange interactions. 
It is therefore necessary to consider an additional energy gain associated with the formation of the CSL state in magnetic fields, analogous to the energy gain that stabilizes a magnetic skyrmion lattice against antisymmetric exchange interactions \cite{Matsumura24}.

\subsection{atomic displacement in magnetic fields}
The atomic displacement occurs with the same periodicity as the conical magnetic structure in magnetic fields applied along the $c$ axis. 
From symmetry considerations, this is a natural result as observed also in the spin-lattice coupling case \cite{Arima07}. 
Here in \YbNiCuAl\ of $f$-electron system with nonzero orbital moment, field induced quadrupole moment accompanying with the conical structure is likely to be the origin of the atomic displacement. 
When the field is applied along $\hat{\mib{z}}$, the ordered magnetic moment along $\hat{\mib{x}}$ tilts toward $\hat{\mib{z}}$, giving rise to an $O_{zx}$ quadrupole moment. 
In the conical structure, this $O_{zx}$ quadrupole rotate about the $z$ axis simultaneously with the magnetic moment. 
The atomic displacement observed here in the experiment is considered to be due to the quadrupole-strain coupling. 

\begin{figure}[t]
\begin{center}
\includegraphics[width=5.5cm]{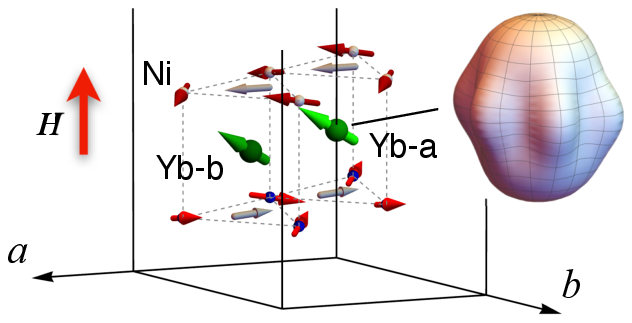}
\end{center}
\caption{(Color online) 
Schematic illustration of the conical magnetic ordered state for $H \parallel c$ of \YbNiAl. 
The arrows on the Yb atoms represent the magnetic moments of Yb. 
The arrows on the Ni sites indicate a possible model of atomic displacements in magnetic fields, which are allowed by the irreducible representation for $\mib{q}=(0, 0, 0.82)$. The arrows at the center of the Ni triangle show the vector sum of the three Ni displacements. 
The calculated charge distribution for Yb-a is also shown, with its asphericity emphasized. 
}
\label{fig:NishiftModel}
\end{figure}

It is important to note that no atomic displacements are detected at zero field. 
In the helical structure at zero field, displacements of surrounding Ni and Al atoms are allowed through the quadrupole-strain coupling. 
The quadrupole moment associated with the magnetic moment lying in the $ab$ plane is either $O_{22}$ or $O_{xy}$. 
However, in the CEF states expressed by $|\pm 7/2\rangle$ and $|\mp 5/2\rangle$, the matrix elements of $O_{22}$ and $O_{xy}$ are all zero within the $4\times 4$ subspace spanned by the ground and first excited doublets. 
This is likely the reason for the absence of lattice distortion at zero field. 
On the other hand, the matrix elements of $O_{zx}$ and $O_{yz}$ have finite values between the ground and first excited doublets. 
Therefore, in magnetic fields, when the two doublets are mixed, the $O_{zx}$ and $O_{yz}$ quadrupole moments are linearly induced by the applied field, which is consistent with the field dependence of $F_{\text{C}}$ shown in Fig.~\ref{fig:PRthscans0}(e).
A calculated charge distribution for a tilted magnetic moment is shown in Fig.~\ref{fig:NishiftModel}. 

With respect to the atomic displacements, they should be consistent with the irreducible representation of the conical state. 
The atomic displacements of the Ni atoms shown in Fig.~\ref{fig:NishiftModel} represent one such example allowed by the trigonal symmetry. 
The total displacement of the Ni atoms in the layers above and below each Yb atom reflects the deformation of the $4f$ charge density and the appearance of $O_{yz}$ and $O_{zx}$ quadrupole moments. 
Although not shown, similar atomic shifts are expected to occur for the Al atoms so as to remain consistent with the Yb quadrupole moments. 

\t should be noted that, in the above discussions, we have neglected modifications of the exchange interaction induced by atomic displacements. 
Such effects are explicitly considered in analyses of magnetoelastic effects in incommensurate magnetic orderings in elemental rare-earth metals~\cite{Iwasa94,Ohsumi98,Ohsumi02}. 
Atomic displacements of the Ni, Al, and Yb atoms modify the exchange energy in addition to the CEF energy, and should therefore be treated in a single Hamiltonian. 
However, modeling such effect is beyond the scope of the present paper.

\section{Summary}
The conical magnetic ordered states of the uniaxial chiral helimagnet \YbNiCuAl\ have been studied by resonant X-ray diffraction in magnetic fields applied along the helical $c$ axis. 
The magnetic diffraction peaks clearly disappear at the critical field corresponding to the transition from the conical to the field-induced ferromagnetic state. 
The critical fields were determined to be 4 T and 7 T for $x=0$ and $x= 0.05$, respectively, which had been hardly discernible in the magnetization measurements. 
Simultaneously with the onset of conical order, atomic displacements occur with the same wavenumber as the conical modulation, likely reflecting the coupling between the field-induced quadrupole moments and the lattice. 

We also discussed the transition temperature $T_{\text{N}}$ and the critical fields for $H \parallel c$ ($H_{\text{c}}^{z}$) and $H\perp c$ ($H_{\text{c}}^{x}$) based on a mean-field calculation for a simple $\mib{q}=(0, 0, 1)$ 120$^{\circ}$ model. 
We showed that $T_{\text{N}}$ and $H_{\text{c}}^{z}$ primarily reflect the dominant intralayer exchange interactions within a Yb layer, whereas $H_{\text{c}}^{x}$ reflects the much weaker interlayer interactions. 

\vspace{3mm}
\small{
\paragraph{\small{Acknowledgments}}
We wish to express our sincere gratitude to Y. Tanaka and H. Nakao for generous support during the experiment at the RIKEN beamline BL19LXU of SPring-8 and at BL-3A of the Photon Factory, respectively. 
We also acknowledge valuable discussions with Y. Kato and J. Kishine. 
This work was supported by JSPS Grant-in-Aid for Scientific Research (Nos. JP18K03539, JP20H01854, JP21K13879, and JP21K03467) and JSPS Grant-in-Aid for Transformative Research Areas (Asymmetric Quantum Matters, Nos. JP23H04867 and JP23H04870). 
The synchrotron radiation experiments were performed at BL19LXU of SPring-8 with the approval of RIKEN (Proposal No. 20200083) and at BL-3A of the Photon Factory, KEK, with the approval of the Photon Factory Program Advisory Committee (Nos. 2022G114 and 2024S2-002). 
MT was supported by JST, the establishment of university fellowships toward the creation of science technology innovation (Grant No. JPMJFS2129).
}

\appendix
\section{Crystal chirality}
\begin{figure}[t]
\begin{center}
\includegraphics[width=6cm]{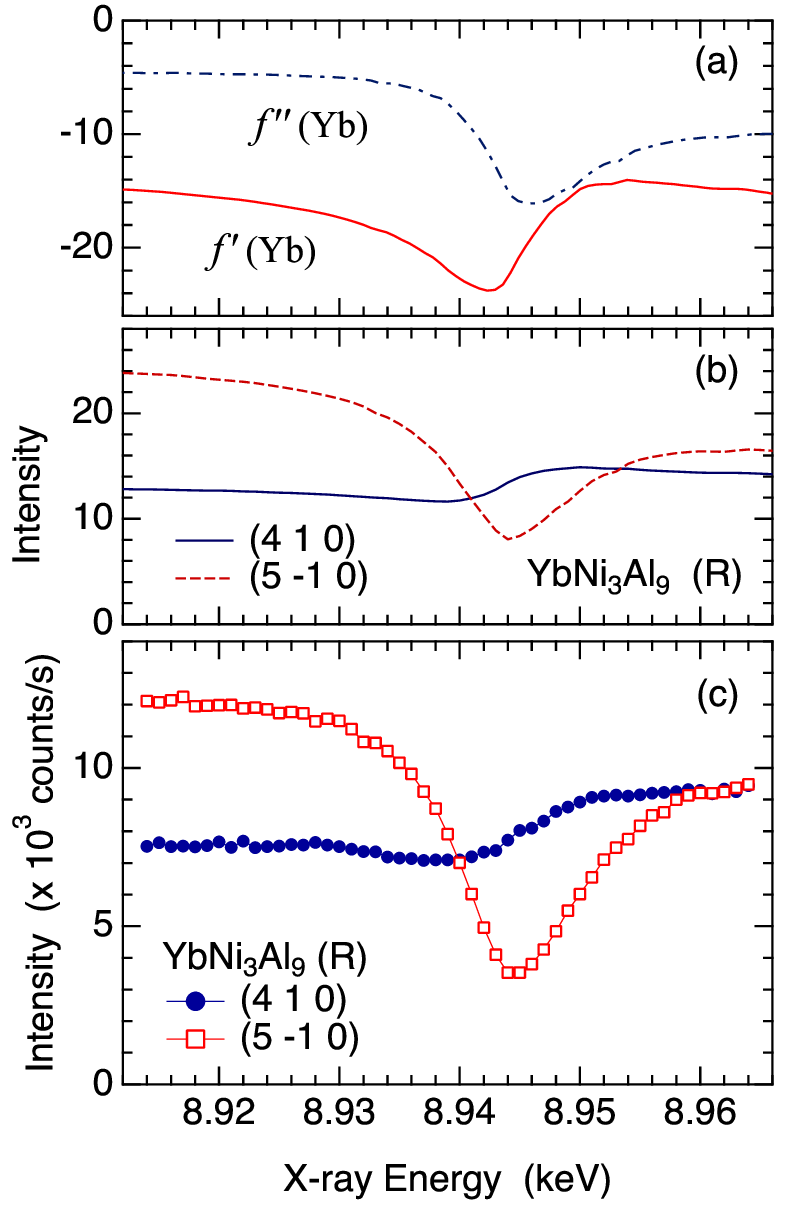}
\end{center}
\caption{(Color online) 
(a) Real and imaginary parts of the atomic scattering factor of Yb, deduced from the absorption coefficient of \YbNiAl\ shown in Fig.~\ref{fig:Escan0T}(b). 
(b) Calculated X-ray energy dependence of the intensities of the (4 1 0)  and (5 $\bar{1}$ 0) fundamental Bragg reflections, assuming the right-handed crystal structure. 
(c) Observed energy dependences of the (4 1 0)  and (5 $\bar{1}$ 0) fundamental Bragg reflections. 
}
\label{fig:Escan410R}
\end{figure}

Figure \ref{fig:Escan410R}(a) shows the real and imaginary parts of the atomic scattering factor of Yb, obtained from the absorption coefficient shown in Fig.~\ref{fig:Escan0T}(b). 
Using the relation $\mu/\rho = -2r_{\text{e}}\lambda \sum_{j}f_j^{\prime\prime}(\omega)/M$, we converted $\mu$ into the imaginary part of the atomic scattering factor $f^{\prime\prime}$ of Yb, from which the real part $f^{\prime}$ was obtained by Kramers-Kronig transformation. 
These scattering factors near the absorption edge were used to calculate the energy dependences of the (4 1 0) and (5 $\bar{1}$ 0) fundamental Bragg reflections and to compare them with the observations.

The crystal chirality of the \YbNiAl\ sample was determined by measuring the energy dependence of an appropriate pair of fundamental Bragg peaks, as shown in Fig.~\ref{fig:Escan410R}(c). 
The structure factor of the (4 1 0) fundamental reflection for the right-handed crystal is expressed as 
$F_{\text{R, (4 1 0)}}(\omega)=A_{\text{Al}} f_{\text{Al}}(\omega)+A_{\text{Ni}}f_{\text{Ni}}(\omega)+A_{\text{Yb}} f_{\text{Yb}}(\omega)$, 
where $A_{\text{Al}}=9.72 - 8.33 i$, $A_{\text{Ni}}=-8.89+15.4 i$, and $A_{\text{Yb}}=6.0$. 
For $F_{\text{R, (5 $\bar{1}$ 0)}}$, the complex conjugates for $A_{\text{Al}}$ and $A_{\text{Ni}}$ are used. 
When $f_{\text{Yb}}(\omega)=f_{0,\text{Yb}} + f_{\text{Yb}}'(\omega) +  if_{\text{Yb}}''(\omega)$ exhibits an anomalous dispersion near the absorption edge, the intensity of the Bragg reflection also exhibits a strong energy dependence. 
Since the intensities of the (4 1 0) and (5 $\bar{1}$ 0) reflections for the right-handed crystal are given by $|F_{\text{R, (4 1 0)}}(\omega)|^2$ and $|F_{\text{R, (5 $\bar{1}$ 0)}}(\omega)|^2$, respectively, they exhibit different energy dependences, as shown in Fig.~\ref{fig:Escan410R}(b), due to the different interference effects. 
The relative intensities are consistent with the assumption that the crystal is right-handed.

\section{Magnetic-field reversal}
\begin{figure}[t]
\begin{center}
\includegraphics[width=6cm]{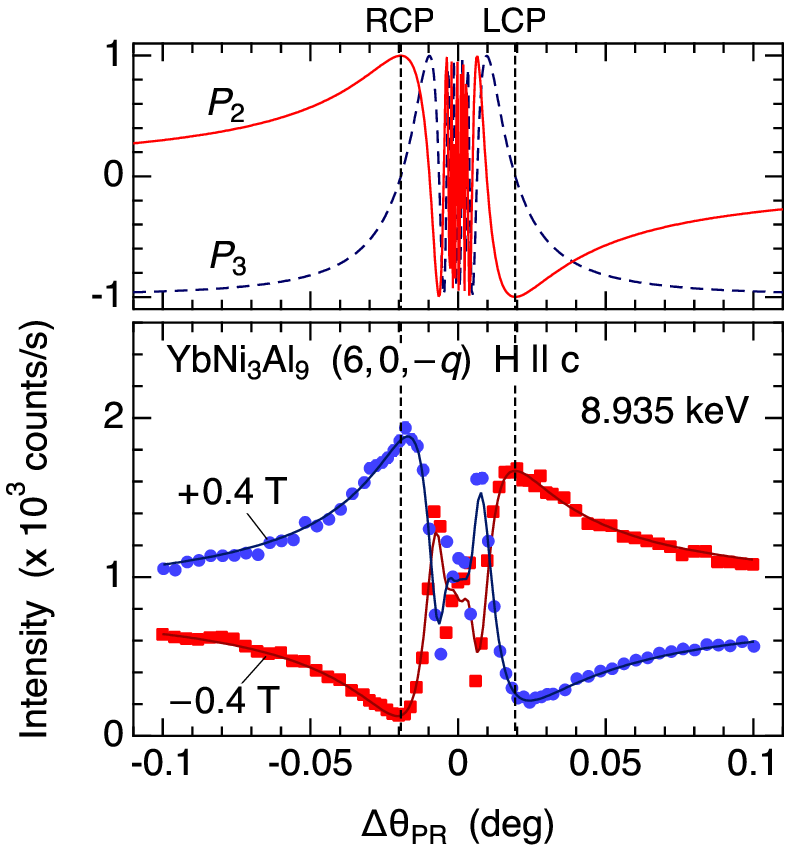}
\end{center}
\caption{(Color online) 
(top) $\Delta\theta_{\text{PR}}$ dependence of the incident polarization state in this experiment, expressed in terms of the Stokes parameters $P_2$ and $P_3$. 
(bottom) $\Delta\theta_{\text{PR}}$ dependence of the peak intensity measured in reversed magnetic fields of $\pm 0.4$ T. 
The X-ray energy was tuned to the $E2$ resonance. Solid lines represent fits using Eq. (\ref{eq:CrossSec2}). 
}
\label{fig:PRthscanE2}
\end{figure}

When the direction of the magnetic field is reversed, the asymmetric behavior with respect to $\Delta\theta_{\text{PR}}$ is also reversed, as demonstrated in Fig.~\ref{fig:PRthscanE2}. Although this measurement was performed at the $E2$ resonance energy, essentially the same result is obtained at the $E1$ resonance. 
From the model given in Eq. (\ref{eq:FMmodel}), there are two possible explanations for this field-reversal effect: (i) the sign of $F_{\text{C}}$ is reversed, or (ii) the magnetic helicity is reversed to $(\hat{\mib{x}} - \hat{\mib{y}})$. 
The latter possibility of helicity reversal, however, is unlikely. 
As shown in Fig.~\ref{fig:NishiftModel}, when the $c$-axis component of the Yb magnetic moment becomes negative in a downward magnetic field, the signs of $O_{yz}$ and $O_{zx}$ quadrupole moments are also reversed, which would lead to atomic displacements of surrounding atoms in opposite directions. 
This naturally explains the experimental observations. 

The top panel of Fig.~\ref{fig:PRthscanE2} shows the $\Delta\theta_{\text{PR}}$ dependence of the incident polarization state in this experiment. 
It should be noted that the sign of $P_2$ is opposite to that in Ref.~\citen{Matsumura17}, because the experimental configuration of the diamond phase plate is different. 
When $P_3=1$ or $-1$, the incident X-ray is linearly polarized parallel ($\sigma$ polarization) or perpendicular ($\pi$ polarization), respectively, to the horizontal scattering plane. 

\section{Contribution of $E2$ resonance at zero field}
\begin{figure}[t]
\begin{center}
\includegraphics[width=6cm]{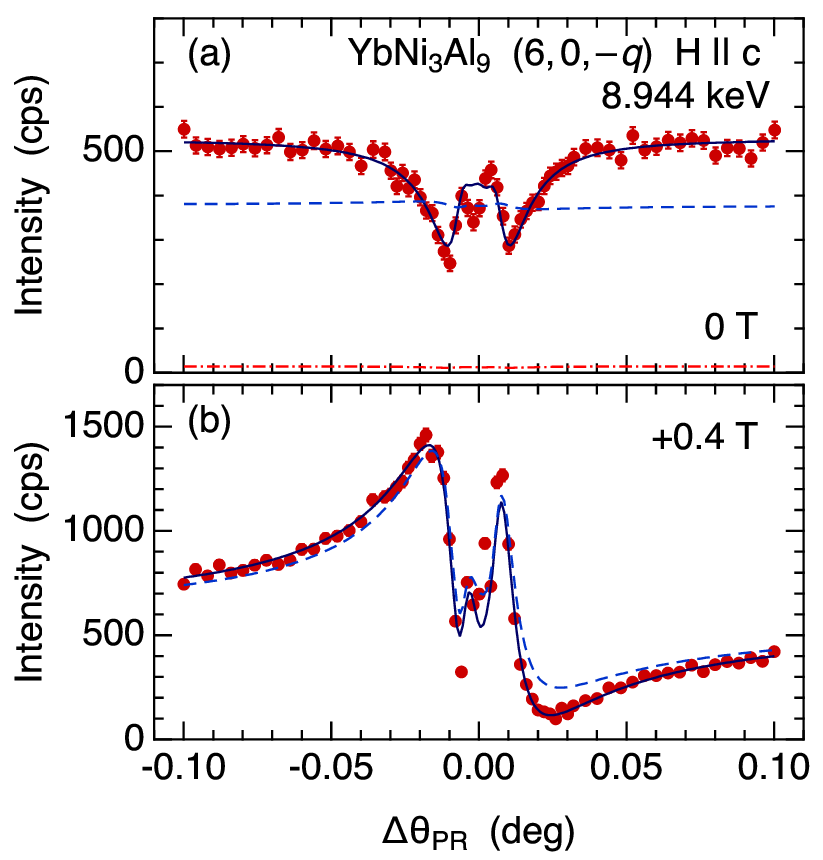}
\end{center}
\caption{(Color online) 
$\Delta\theta_{\text{PR}}$ dependence of the peak intensity measured at (a) 0 T and (b) +0.4 T. 
Solid lines represent fits using Eq. (\ref{eq:CrossSec2}). Dashed lines show calculations considering only the contribution from the $E1$ resonance. The dot-dashed line shows the calculated intensity arising solely from the $E2$ contribution. 
}
\label{fig:PRthscanE1}
\end{figure}

As described in the main text, the $\Delta\theta_{\text{PR}}$ dependence at zero field shown in Fig.~\ref{fig:PRthscans0}(a) cannot be explained by the $E1$ resonance alone. 
To reproduce $C_3/C_0=-0.329$ at zero field, corresponding to the solid line in Fig.~\ref{fig:PRthscans0}(a), it is necessary to take into account interference with the tail of the $E2$ resonance at 8.935 keV, which originates from the $T_{1u}$-type magnetic octupole~\cite{Matsumura17}. 
The contribution of the $E2$ amplitude is approximately one-fifth of that of the $E1$ amplitude, as is estimated from the geometrical structure factor for the $T_{1u}$-type magnetic octupole~\cite{Nagao06}. 
If the interference is neglected and only the $E1$ resonance is considered, the $\Delta\theta_{\text{PR}}$ dependence becomes almost flat as shown by the dashed line in Fig.~\ref{fig:PRthscanE1}(a), where a tiny anomaly reflecting the $P_2$ term appears due to the nonzero value of $q$. 
The intensity of the $E2$ resonance alone is about one twenty-fifth of that of the $E1$ resonance, as shown by the dot-dashed line. 
The double-dip structure therefore arises from the interference effect.

In finite magnetic fields, however, Thomson scattering from lattice distortions dominates the scattering intensity, and the $E2$ contribution can be neglected. 
As shown by the dashed line in Fig.~\ref{fig:PRthscanE1}(b) for 0.4 T, the model structure factor given by Eq. (\ref{eq:CrossSec2}) reproduces the experimental results well.

\bibliographystyle{jpsj}
\bibliography{71770}

\end{document}